\documentclass{Interspeech2024}
\usepackage{float} 
\usepackage[font=small,skip=0pt]{caption}

\interspeechcameraready

\title{Comparative Analysis of Personalized Voice Activity Detection Systems: Assessing Real-World Effectiveness}

\name[affiliation={1,*,\dagger}]{Satyam}{Kumar}
\name[affiliation={2,*}]{Sai Srujana}{Buddi}
\name[affiliation={2}]{Utkarsh Oggy}{Sarawgi}
\name[affiliation={2}]{Vineet}{Garg}
\name[affiliation={2}]{Shivesh}{Ranjan}
\name[affiliation={2}]{Ognjen (Oggi)}{Rudovic}
\name[affiliation={2}]{Ahmed}{Hussen Abdelaziz}
\name[affiliation={2}]{Saurabh}{Adya}

\address{
  $^1$The University of Texas at Austin, USA 
  $^2$Apple, USA
}

\email{satyam.kumar@utexas.edu, sbuddi@apple.com}


\begin{document}
\maketitle

\begin{abstract}
Voice activity detection (VAD) is a critical component in various applications such as speech recognition, speech enhancement, and hands-free communication systems. With the increasing demand for personalized and context-aware technologies, the need for effective personalized VAD systems has become paramount. In this paper, we present a comparative analysis of Personalized Voice Activity Detection (PVAD) systems to assess their real-world effectiveness. We introduce a comprehensive approach to assess PVAD systems, incorporating various performance metrics such as frame-level and utterance-level error rates, detection latency and accuracy, alongside user-level analysis. Through extensive experimentation and evaluation, we provide a thorough understanding of the strengths and limitations of various PVAD variants. This paper advances the understanding of PVAD technology by offering insights into its efficacy and viability in practical applications using a comprehensive set of metrics.
\end{abstract}

\def\thefootnote{*}\footnotetext{These authors contributed equally to this work}\def\thefootnote{\arabic{footnote}}
\def\thefootnote{\fnsymbol{footnote}}
\footnotetext[2]{Work done at Apple}
\def\thefootnote{\arabic{footnote}}

\section{Introduction}


VAD systems are crucial for modern speech recognition pipelines, allowing for the identification of speech and non-speech segments on a frame-by-frame basis and initiating downstream tasks such as speech recognition, enhancement, and endpointing \cite{sohn1999statistical, ramirez2004efficient, ramirez2007voice, oggy2023streaming, buddi2023efficient, chang2019joint}. However, in real-world scenarios involving multiple users, traditional VAD systems often trigger erroneously with disruptive false positives, necessitating Personalized Voice Activity Detection (PVAD) systems \cite{chen2020end}.

Traditional cascaded VAD and speaker verification models can address the PVAD task but fall short of meeting comprehensive requirements for seamless integration. PVAD systems, commonly used alongside speech recognition and other models, demand lightweight solutions for widespread real-time deployment \cite{buddi2023efficient}. While standalone speaker verification systems are reliable, they come with high resource demands, operational costs, and detection latencies \cite{variani2014deep}. With the imperative for real-time processing and triggering downstream tasks, low detection latency becomes crucial for PVAD systems. Therefore, PVAD systems must effectively balance accuracy, efficiency, and responsiveness to cater to various applications.

The introduction of PVAD by Ding et al. \cite{ding2019personal} represented a significant advancement, offering a lightweight solution for robust target speaker speech detection. Unlike conventional methods relying on heavy speaker verification models generating dynamic speaker information in real-time, this approach extracts speaker information from enrollment utterances beforehand and integrates this static speaker information into the VAD architecture. In \cite{ding2022personal}, authors extended the PVAD architecture and assessed its effectiveness in downstream Automatic Speech Recognition (ASR) tasks. Addressing the challenge of missing user-specific enrollments, Makishima et al. \cite{makishima2021enrollment} proposed an approach for enrollment less training of PVAD. Cheng et al. \cite{cheng2023target}  explored target speaker detection in the context of diarization. Medennikov et al \cite{medennikov2020target} proposed techniques rooted in dinner party scenarios involving multi-speaker diarization.  Other recent works have proposed competitive methods for target speaker voice activity detection \cite{he2021target, wang2022cross, kang2023svvad, wang2023target,tan2020rvad}.

While these studies have introduced different forms of personalization in the voice activity detection pipeline, they often lack systematic evaluations across a range of crucial real-world metrics of accuracy, efficiency, and responsiveness. Notably, none of the studies have compared or benchmarked responsiveness, a critical factor in real-time systems. Furthermore, in personalized machine learning, it is essential to systematically assess any approach to ensure consistent performance improvements across users \cite{rudovic2018personalized}, lacking in previous PVAD works.

Finally, for real-world deployment, PVAD systems must adapt to a wide range of devices and environments. While high-end devices support intricate PVAD models, resource-constrained devices like smartwatches require lightweight and efficient solutions \cite{buddi2023efficient}. To this end, our paper presents a detailed analysis of various fusion strategies for PVAD systems. By examining these strategies across different performance metrics, we aim to provide insights into their efficacy across diverse devices and usage scenarios, ultimately contributing to the development of versatile and optimized PVAD systems rooted in real-world viability. We first establish the components of PVAD systems and a variety of its comparative fusion strategies (Section  \ref{sec:systems}), then introduce the evaluation metrics and experimental setup (Sections \ref{sec:metrics} and \ref{sec:exp_setup}), and finally discuss the results and conclusion (Sections \ref{sec:results_and_discussion} and \ref{sec:conclude}).

\begin{figure*}
    \centering
    \includegraphics[width=\linewidth,height=0.36\linewidth]{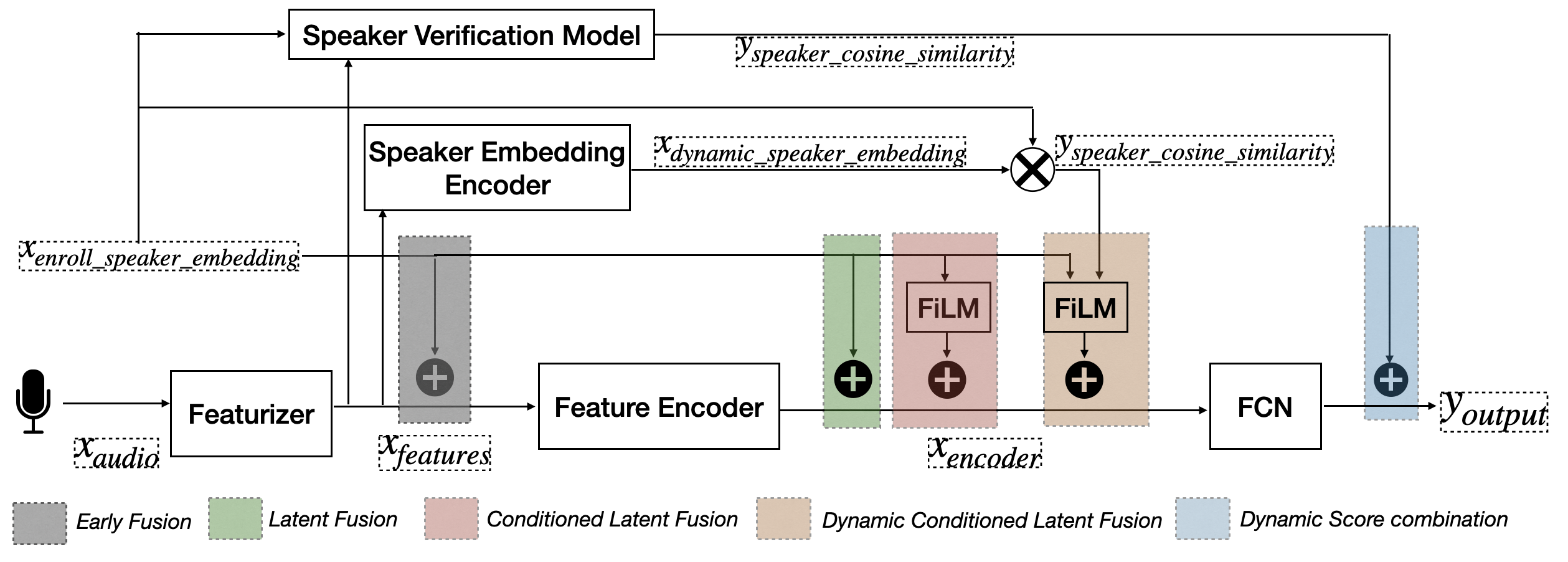}
    \caption{ Pictorial depiction of different fusion strategies for personalization in  Voice Activity Detection Systems.}
        \label{fig:graphicalAbstract}
\end{figure*}

\section{System and Components}\label{sec:systems}
A typical PVAD system includes VAD for speech detection, a speaker verification model, and fusion to integrate information from both components to detect the presence of the target speaker. For speaker verification, we utilized a pre-trained d-vector model \cite{wan2018generalized} to generate voice characteristic embeddings referred to as ``d-vectors". These d-vectors, computed and stored during a one-time enrollment process from user recordings, facilitate speaker verification using cosine similarity \cite{george2015cosine} and are denoted as $y_{speaker\_cosine\_similarity}$. Conventional VAD systems with mel filter bank features and LSTM networks \cite{7472768,sak2014long} were employed to develop an end-to-end LSTM-based network for frame-level binary classification of `speech' ($p_s$) and `no speech' ($p_{ns}$) at each frame ($y_i$). The fusion component processes VAD and speaker verification component outputs to generate $y\in{p_{ts}, p_{nts}, p_{ns}}$ per frame, where $p_{ts}$ denotes target speaker speech, $p_{nts}$ denotes non-target speaker speech, and $p_{ns}$ denotes no speech. Following sections discuss approaches to achieve this fusion.


\subsection{Fusion Strategies for PVAD} \label{ssec:pvad_fusion}
In this section, we explore various PVAD variants that integrate Speaker Verification and VAD systems using multimodal fusion techniques \cite{baltruvsaitis2018multimodal}. These variants, including early, latent, and score fusions, utilize both static and dynamic speaker embeddings. While multiple combinations are feasible, we concentrate on established architectures shown in Figure \ref{fig:graphicalAbstract}. Unless noted otherwise, all methods rely on static speaker information.

\subsubsection{Dynamic Score Combination (DSC)}\label{ssec:sc}
In this approach, acoustic features undergo processing by VAD and Speaker verification models as discussed earlier, resulting in speech detection and speaker verification scores per frame. These scores are then integrated to produce PVAD outputs per frame ($f^{i}$). However, operating at the frame level for both VAD and speaker verification models, this method is resource-intensive. 

\subsubsection{Early Fusion (EF)}\label{sssec:ef}
Drawing inspiration from the approach described in \cite{ding2019personal}, we combine the acoustic features $x_{features}$ with the enrollment speaker embeddings $x_{enroll\_speaker\_embedding}$ for every frame $f_i$ across all utterances in the dataset. This combined feature set is utilized to train a PVAD model end-to-end, which predicts the likelihood of the target speaker's speech presence for each frame. As this approach relies on static enrollment speaker information, it eliminates the necessity of running speaker verification models, thus enhancing resource efficiency.

\subsubsection{Latent Fusion (LF)}\label{sssec:lf}
The EF technique discussed in Section \ref{sssec:ef} yields high-dimensional feature spaces, leading to a parameter-intensive model. To address this, we employ a two-step process. Initially, a VAD-like network extracts speech embeddings, $x_{encoder}$, from acoustic features, $x_{features}$, capturing crucial information about speech presence. Subsequently, we augment these speech embeddings with user-specific embeddings, $x_{enroll\_speaker\_embedding}$, creating a unified representation fed to Fully Connected Network (FCN) of PVAD. This approach effectively manages input feature dimensionality while preserving the advantages of static early fusion as discussed earlier.

\subsubsection{Conditioned Latent Fusion (CLF)}\label{sssec:CF}
Combining speaker embeddings and acoustic embeddings, as discussed in \ref{sssec:lf}, is suboptimal because they capture different information modalities, which restricts the model's learning capabilities without suitable inductive biases. To remedy this, we utilize Feature-wise Linear Modulation (FiLM) modulation \cite{perez2018film}, inspired by \cite{ding2022personal}, wherein speaker embeddings modulate the acoustic embeddings of each frame, thereby enhancing the fusion of multimodal information.

\subsubsection{Dynamic Conditioned Latent Fusion (DCLF)}\label{sssec:DCLF}
In this strategy, we improve CLF by incorporating dynamic speaker information through speaker embedding extraction using a lightweight speaker embedding encoder network along with static enrollment embeddings. Following \cite{ding2022personal}, we jointly train this network with the conditioned fusion architecture, integrating the enrollment speaker embedding and cosine similarity between enrollment and frame-level dynamic speaker embeddings from the lightweight extractor (Figure \ref{fig:graphicalAbstract}).

\section{Evaluation Metrics}\label{sec:metrics}
Considering the diverse applications of PVAD in triggering downstream tasks such as speech recognition, diarization, and endpointing, a well-evaluated PVAD model has the potential to be versatile across multiple use cases. To align with different usage scenarios and requirements, we have employed the following metrics to evaluate PVAD systems:

\subsection{Equal Error Rate}
Equal Error Rate (EER) represents the point on the Detection Error Tradeoff (DET) curve \cite{martin1997det} where the False Positive Rate (FPR) equals the False Negative Rate (FNR). A lower EER is preferable. We employed the following EER metrics to determine system accuracy owing to real-world usage:

\begin{itemize}
    \item Frame-level EER (fEER)
        \begin{itemize}
            \item In the PVAD setting, fEER PVAD measures the system's accuracy in distinguishing between targeted user speech, non-targeted user speech, and no speech at the frame-level.
            \item In VAD setting, fEER VAD measures the system's accuracy in distinguishing between speech and no speech at frame-level where enrollment speaker embeddings are missing\cite{7015406}.
        \end{itemize}
    \item Utterance-level EER (uEER)
        \begin{itemize}
            \item In PVAD setting. uEER PVAD aggregates granular frame-level understanding to the utterance-level, providing a better representation of the system's effectiveness in triggering downstream tasks. 
        \end{itemize}
\end{itemize}

\subsection{Detection Latency}
While EER is crucial for model assessment and comparison, detection latency in targeted speech detection is pivotal for real-time responsiveness, particularly in PVAD systems triggering downstream tasks real-time. In this study, we quantify detection latency at the utterance-level, defined as the time from targeted speech onset to its detection at the designated uEER operating point.

\subsection{Detection Accuracy}
To evaluate improvements in target speaker detection per user, we determine the accuracy of detecting the target speaker in their test utterances at the uEER operating point. Detection accuracy is calculated as the percentage of utterances where the target user's speech is accurately detected out of the total utterances spoken by the target user.

\section{Experimental Setup} \label{sec:exp_setup}
\subsection{Dataset}
We experimented with the LibriSpeech dataset \cite{panayotov2015librispeech}, containing approximately $960$ hours of speech from $2338$ users. The dataset includes meticulously segmented and aligned speech segments from audiobooks, with a training subset of $460$ hours of clean speech and $500$ hours of noisy speech. Test and validation sets each consists of $11$ hours of clean and noisy data.

To simulate natural conversational dynamics, we created concatenated audio files by uniformly sampling segments from $1-3$ users and augmented them with noise from the MUSAN corpus \cite{snyder2015musan}, varying the SNR from $0-30$ dB for increased robustness. Each resulting audio file represented an ``utterance", with one user randomly assigned as the target user.

$3-5$ speech segments are randomly selected from each user as ``enrollment segments" to acquire their ``enrollment speaker embeddings". These embeddings are derived from a speaker verification model and averaged to generate a robust d-vector embedding estimate. To accommodate scenarios where users opt not to enroll their voices, we randomly excluded enrollment speaker embeddings for $20\%$ of concatenated utterances. In such cases, target enrollment embeddings were replaced with $256-$dimensional zero embeddings, treating both target and non-target speakers as the target speaker.

In each utterance, segments corresponding to the target speaker were labeled as $ts$, segments from other speakers as $nts$, and non-speech segments as $ns$, facilitating differentiation of speech segments based on speaker identity and content for PVAD model training and evaluation.

\subsection{Model Architecture}
The featurizer (shown in Figure \ref{fig:graphicalAbstract}) extracts, $x_{features}$, a $40$-dimensional log-mel filter bank features per frame, employing a frame length of $25$ms and a step size of $10$ms, resulting in features at $100$Hz for $x_{audio}$ in the dataset. The feature encoder, depicted in Figure \ref{fig:graphicalAbstract}, consists of a $2$-layer LSTM network with a hidden size of $64$, capturing temporal dependencies within utterances. This is followed by two FCN layers with tanh activation functions to enable nonlinear transformations of the extracted embeddings. In the CLF strategy, we employed a FiLM layer to modulate the speech embeddings produced by the feature encoder using  the speaker embedding, before passing it to FCN layers. As for DCLF, we utilized a single-layer LSTM (Figure \ref{fig:graphicalAbstract}: Speaker embedding generator) to generate a dynamic $256-$dimensional speaker embedding, which is then used to extract $y_{speaker\_cosine\_similarity}$ for conditioned fusion.

\subsection{Training Setup}\label{ssec:train_setup}
We trained and evaluated five models: one for score combination in the VAD setting as discussed in Section \ref{ssec:sc}, and one for each end-to-end PVAD variant described in \ref{ssec:pvad_fusion}. All models used cross-entropy loss, binary for VAD and categorical for PVAD, with training employing the Adam optimizer \cite{kingma2014adam} at a learning rate of $1e^{-3}$.

\section{Results and Discussion} \label{sec:results_and_discussion}
We evaluated five models trained as described in Section \ref{ssec:train_setup}, using the evaluation metrics outlined in Section \ref{sec:metrics}. The standard DSC model, detailed in Section \ref{ssec:sc}, serves as our baseline for comparison against the end-to-end trained PVAD models.
\subsection{Accuracy: Equal Error Rate}
\subsubsection{Frame level Analysis}
Table \ref{tab:eer_perf} illustrates the frame-level Equal Error Rate (fEER) comparison among all PVAD variants in both PVAD and VAD settings, with and without enrollment speaker embeddings. Across the PVAD setting, the four end-to-end PVAD variants outperform the DSC baseline. Notably, the CLF variant marginally outperforms DCLF in terms of fEER PVAD, indicating that DCLF, resembling a standalone speaker verification model, requires more frames for precise prediction.
\begin{table}[H]
  \centering
  \begin{tabular}{lccccc}
    \toprule
    & \textbf{DSC} & \textbf{EF} & \textbf{LF} & \textbf{CLF} & \textbf{DCLF} \\
    \midrule
    \textbf{fEER PVAD} & 26.2 & 12.2 & 10.6 & \textbf{\textit{10.2}} & 10.4 \\
    \textbf{fEER VAD} & \textbf{\textit{5.9}} & 7.4 & 6.6 & 6.6 & 6.6 \\
    \textbf{uEER PVAD} & 11.6 & 11.2 & 11.3 & 10.6 & \textbf{\textit{9.2}} \\
    \bottomrule
  \end{tabular}
\caption{EER comparison of PVAD variants.}
\vspace{-1em}
\label{tab:eer_perf}
\end{table}
Ideally, PVAD systems should achieve comparable performance to VAD systems in detecting speech, in scenarios with missing speaker enrollment information. Table \ref{tab:eer_perf} displays the fEER VAD of all PVAD systems evaluated as VAD systems. While the DSC variant shows marginally superior fEER compared to other fusion strategies, likely due to the task-specific training of DSC's VAD component, the rest of the end-to-end PVAD models introduce a new dimension for inferring the presence of the target speaker through lightweight fusion.

\subsubsection{Utterance level Analysis}
We smoothed the frame-level PVAD output scores to extract an utterance-level score by calculating a moving average over a window frame of 5 and selecting the highest score. Table \ref{tab:eer_perf} compares the utterance-level EER  of PVAD variants for the PVAD task. Our findings reveal that all fusion strategies outperform the baseline DSC variant. Notably, CLF outperforms basic concatenation-based fusions like LF and EF due to the FiLM-based fusion strategy. This strategy enhances information integration by leveraging speaker embeddings and acoustic features, resulting in improved CF performance over EF and LF. Additionally, DCLF outperforms other fusion approaches by integrating dynamic speaker cosine similarity scores extracted at the frame-level with the static enrollment speaker embedding. Due to the absence of speech-free utterances in the LibriSpeech dataset, uEER computation for the VAD task was omitted.

\subsection{Detection Latency and Detection Accuracy}\label{ssec:medianaccuracy}
\begin{figure}[h]
    \centering
    \includegraphics[width=\linewidth]{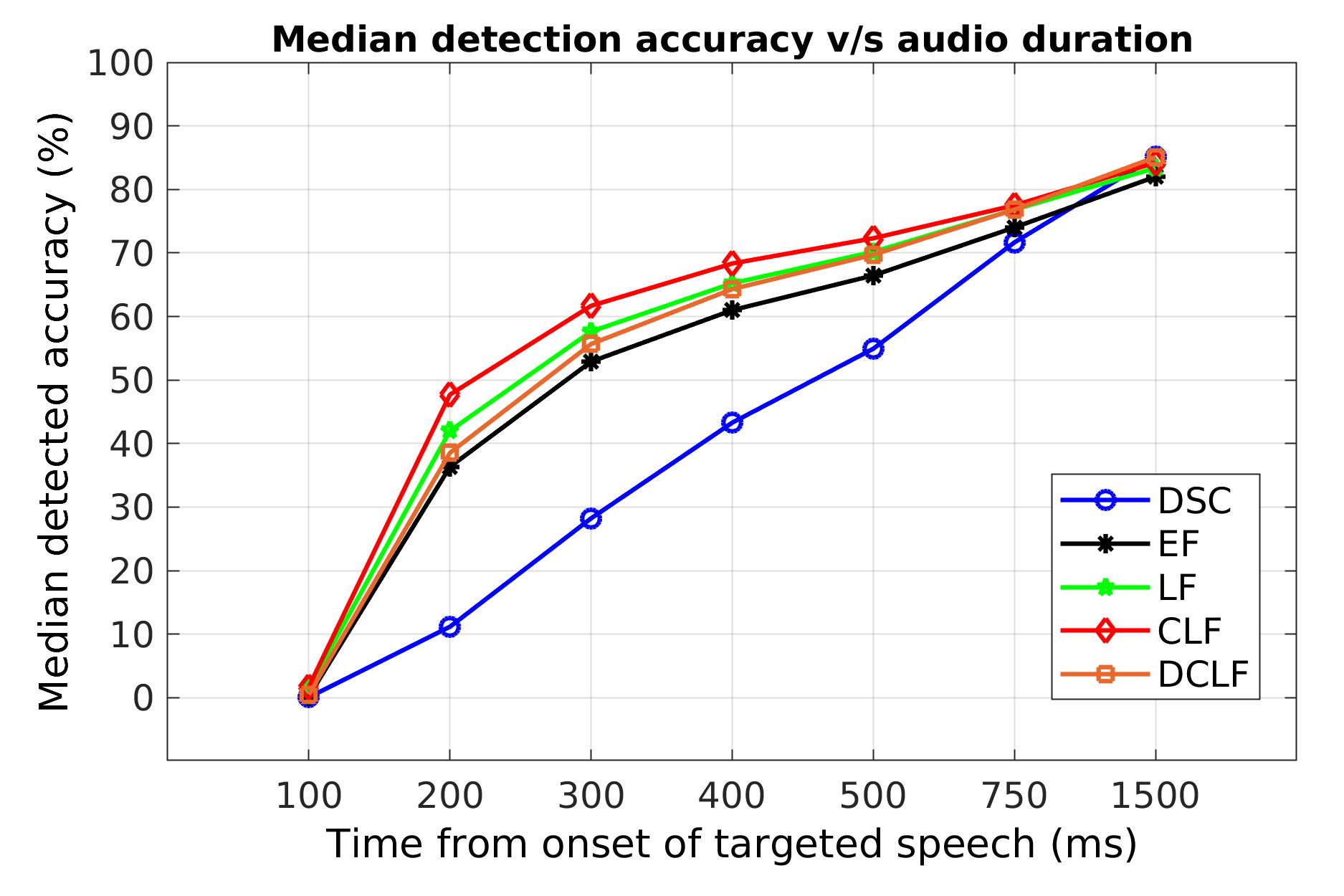}
    \caption{Impact of audio duration on detection accuracy}
\label{fig:acceptance_rate}

\vspace{-1em}
\end{figure}

Studies indicates that speaker verification models achieve higher accuracy with increased audio context \cite{eronen2005audio}. Figure \ref{fig:acceptance_rate} explores the impact of audio duration on PVAD model accuracy at the uEER operating point. End-to-end PVAD variants notably outperform DSC accuracies with shorter audio context, highlighting their high responsiveness and reliability, making them ideal for real-time streaming applications. As audio context increases, their performance converges.

Table \ref{tab:latency} summarizes the median detection latency and accuracy of all users in the test set. It shows that all PVAD variants surpass the baseline DSC in both metrics, achieving a minimum 40\% improvement in latency alongside enhanced accuracy.
\begin{table}[h]
  \centering
  \begin{tabular}{lccccc}
    \toprule
    & \textbf{DSC} & \textbf{EF} & \textbf{LF} & \textbf{CLF} & \textbf{DCLF} \\
    \midrule
    \textbf{Latency} & 432.5 & 240 & 202.5 & \textbf{185} & 240 \\
    \textbf{Accuracy} & 93.2 & 96.7 & 96.6 & 95.9 & \textbf{96.8} \\
    \bottomrule
  \end{tabular}
\caption{Median latency (ms) and accuracy results of PVAD models.}
\label{tab:latency}
\vspace{-2em}

\end{table}

\subsection{User-level evaluation}

\begin{figure}[!ht]
    \centering
    \includegraphics[width=1.05\linewidth]{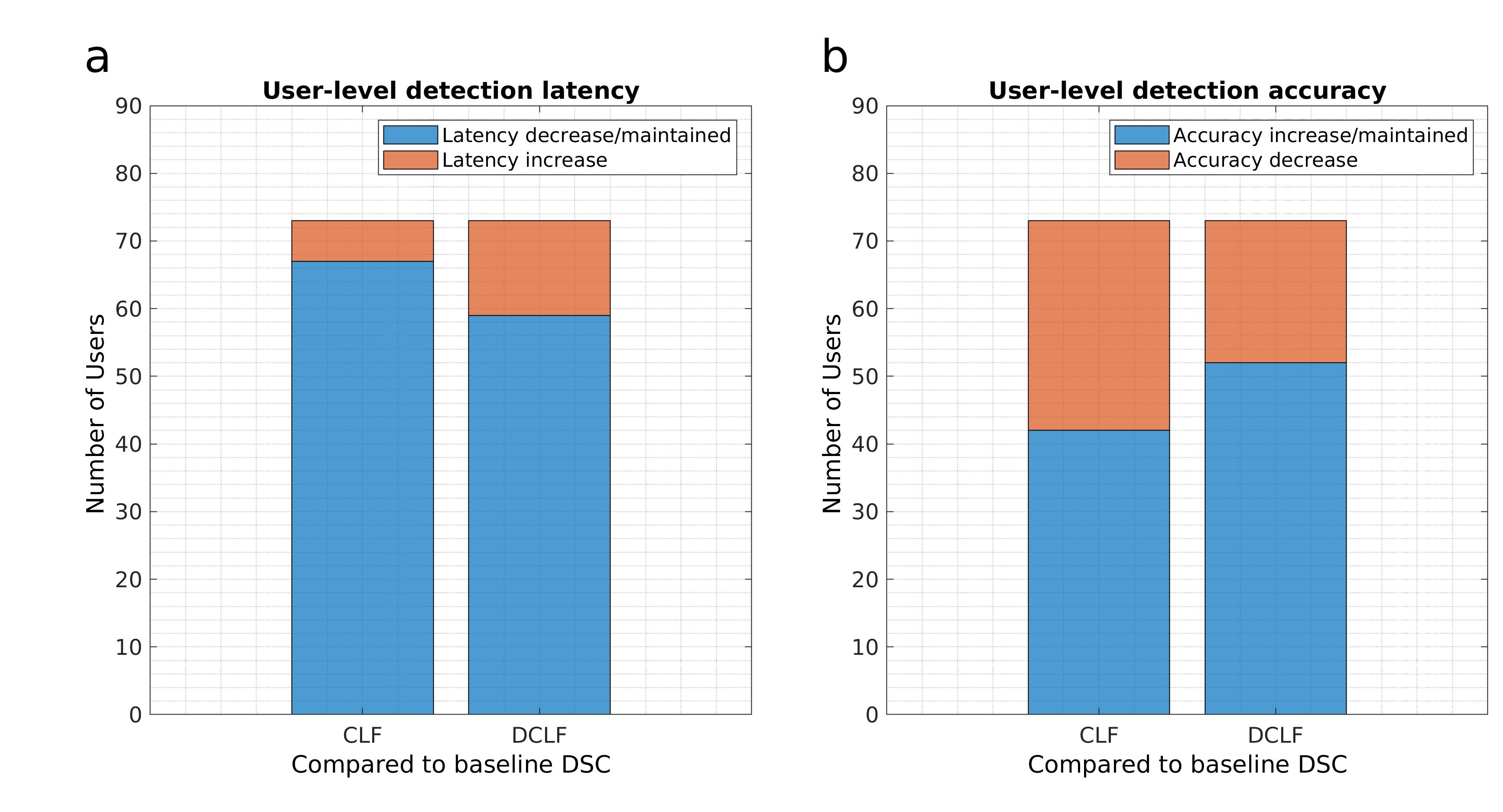}
    \caption{ User-level analysis of  a) detection latency,   b) accuracy of target speaker presence.}
        \label{fig:subjectLevel}
        \vspace{-2em}
\end{figure}

Ensuring consistent performance improvements across users is crucial for personalization\cite{fan2006personalization}. To explore this aspect, we analyze the results discussed in Section \ref{ssec:medianaccuracy} at a user level across 73 users in the test set, focusing on the best-performing PVAD variants, CLF and DCLF, and comparing them against the baseline DSC. Figure \ref{fig:subjectLevel} illustrates a user-level comparison of latency among DSC, CLF, and DCLF. In terms of detection latency, both CLF and DCLF significantly outperform DSC (p-value $< 0.01$ using Wilcoxon signed-rank test \cite{woolson2007wilcoxon}). Precisely, CLF and DCLF improve latency over DSC in 67 and 59 out of 73 users, respectively. Finally, both DCLF and CLF results in increase or retention of detection accuracy at EER operating point among majority of users compared to DSC (CLF $\geq$ DSC: 42/73, DCLF $\geq$ DSC: 52/73), although no significant difference was observed in any pairwise comparisons of detection accuracy.  

\subsection{Model Efficiency}
The end-to-end PVAD models surpass the baseline across all metrics despite being significantly smaller in size. Specifically, CLF and DCLF consist of only 5\% and 26\%, respectively, of the parameters compared to the baseline DSC, as shown in Table \ref{tab:modelsize}. This trait, along with their superior performance, renders them ideal for on-device deployments.

\begin{table}[h]
  \centering
  \begin{tabular}{lccccc}
    \toprule
    & \textbf{DSC} & \textbf{EF} & \textbf{LF} & \textbf{CLF} & \textbf{DCLF} \\
    \midrule
    \textbf{Parameters} & 1.50 & 0.13 & 0.08 & 0.10 & 0.40 \\
    \bottomrule
  \end{tabular}
\caption{Parameter count of PVAD models (in millions).}
\vspace{-1em}
\label{tab:modelsize}
\end{table}

Our analysis reveals that relying solely on one performance metric is inadequate for evaluating and deploying PVAD systems. While CLF shows superior latency and frame-level EER, DCLF performs better in accurately detecting the target speaker at the utterance level. Different fusion-based models suit specific use cases: CLF or EF for devices with minimal latency and limited on-device requirements, and DCLF for devices prioritizing accuracy.

\section{Conclusion} \label{sec:conclude}
This study investigates various PVAD systems, revealing that lightweight end-to-end models, incorporating static speaker information and significantly smaller than parameter-intensive speaker verification-based models, outperform baseline systems. Our analysis indicates these models enhance the speed of detecting target speaker speech by at least 40\%, crucial for real-world deployment. Moreover, dynamically estimating speaker characteristics improves accuracy over static fusion methods. Additionally, these PVAD models demonstrate comparable performance in VAD tasks compared to models trained solely for VAD. These findings highlight the potential of PVAD systems in real-world applications, offering reliable detection of the target speaker while improving speed and accuracy in speech recognition. PVAD systems represent a promising advancement, demonstrating effectiveness and suitability for practical deployment.

\bibliographystyle{IEEEtran}
\bibliography{mybib}

\end{document}